\documentclass[twocolumn,aps,prl,superscriptaddress,amsfonts]{revtex4-1}

\usepackage{amsmath}
\usepackage{graphicx}
\usepackage{bm}
\usepackage{array}
\usepackage{dcolumn}
\usepackage{hepparticles}
\usepackage{heppennames}
\usepackage{hepnicenames}

\begin{document}

\newcommand{\NowYork}{\altaffiliation{Current address: York Univ., Dept.\ of Phys.\ and Astr., Toronto, Ontario M3J 1P3, Canada}}

\newcommand{\Harvard}{\affiliation{Dept.\ of Physics, Harvard University, Cambridge, MA 02138}}
\newcommand{\Juelich}{\affiliation{IKP, Forschungszentrum J\"{u}lich GmbH, 52425 J\"{u}lich, Germany}}
\newcommand{\JuelichZat}{\affiliation{ZAT, Forschungszentrum J\"{u}lich GmbH, 52425 J\"{u}lich, Germany}}
\newcommand{\York}{\affiliation{York University, Department of Physics and Astronomy, Toronto, Ontario M3J 1P3, Canada}}
\newcommand{\Garching}{\affiliation{Max-Planck-Institut f\"{u}r Quantenoptik, Hans-Kopfermann-Strasse 1, 85748 Garching, Germany}}
\newcommand{\Munich}{\affiliation{Ludwig-Maximilians-Universit\"{a}t M\"{u}nchen, Schellingstrasse 4/III, 80799 M\"{u}nchen, Germany}}
\newcommand{\Mainz}{\affiliation{Institut f\"{u}r Physik, Johannes Gutenberg Universit\"{a}t and Helmholtz Institut Mainz, D-55099 Mainz, Germany}}
\newcommand{\Rowland}{\affiliation{Rowland Inst.\ at Harvard, Harvard University, Cambridge, MA 02142}}

\DeclareRobustCommand{\Hbar}{\HepAntiParticle{H}{}{}\xspace}
\DeclareRobustCommand{\H}{\HepParticle{H}{}{}\xspace}

\DeclareRobustCommand{\pbar}{\HepAntiParticle{p}{}{}\xspace}
\DeclareRobustCommand{\p}{\HepParticle{p}{}{}\xspace}

\DeclareRobustCommand{\pos}{\HepParticle{e}{}{+}\xspace}
\DeclareRobustCommand{\e}{\HepParticle{e}{}{-}\xspace}

\newcommand{\BtrapFigure}{
\begin{figure}[htbp!]
\includegraphics*[width=\columnwidth]{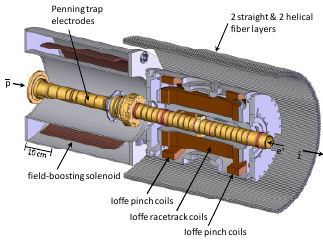}
\caption{Electrodes and coils produce Penning traps (to store \pbar and \pos ) and a Ioffe trap \cite{PritchardProposesIoffeTrap}  (to store \Hbar ). Much of the vacuum enclosure and cooling system is hidden to make the traps and detectors visible. An external solenoid (not shown) adds a 1 T magnetic field along the trap axis $\hat{\bf z}$ which is vertical.}
\label{fig:Btrap}
\end{figure}
}

\newcommand{\TrapAndPotentialsFigure}{
\begin{figure}[htbp!]
\includegraphics*[width=\columnwidth]{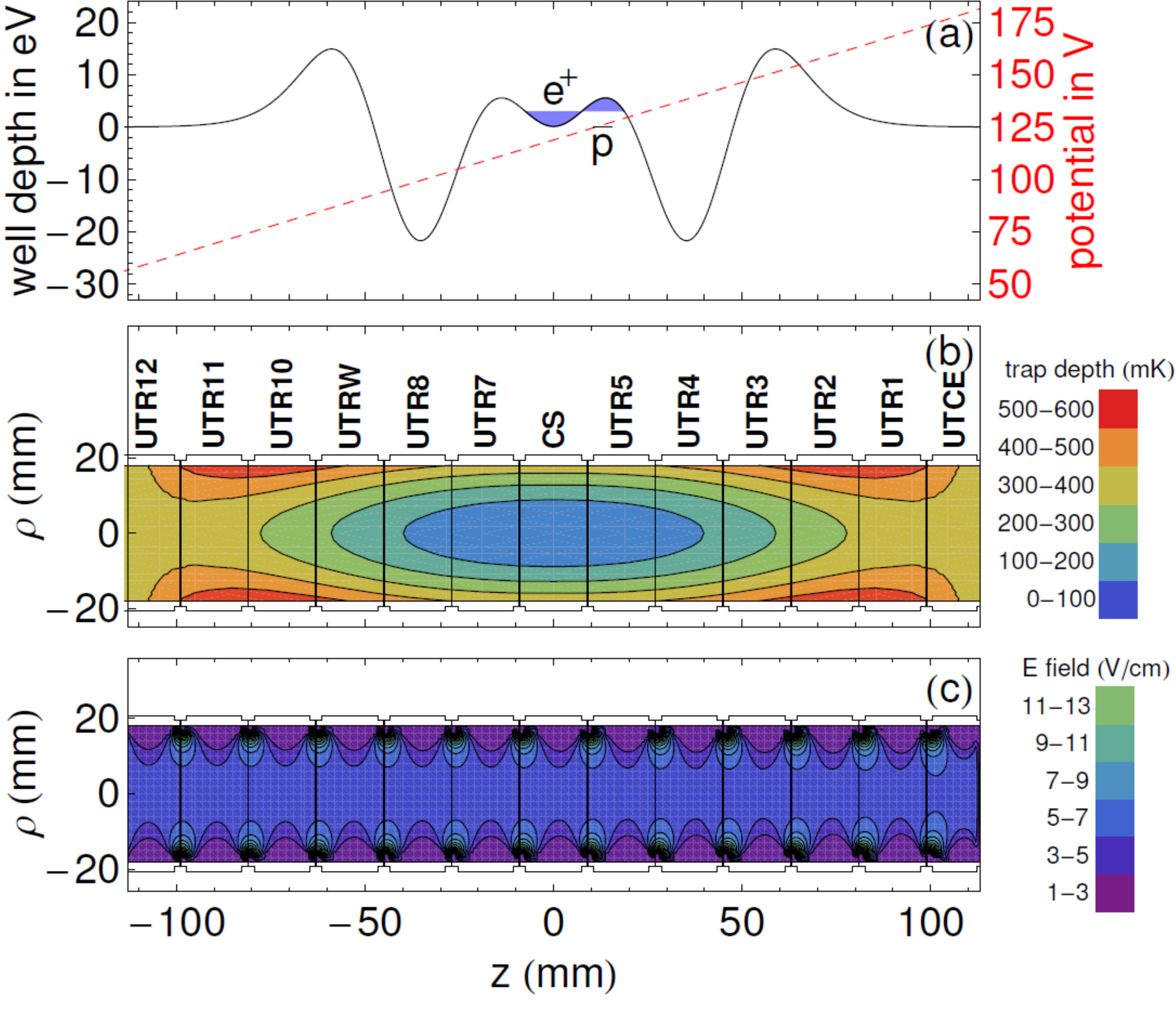}
\caption{(a) Potentials along the center axis used to contain (solid) and remove (dashed) charged particles.  (b) Electrode cross sections with equipotential energy contours for a low-field-seeking, ground-state \Hbar in the Ioffe trap. (c) Axial electric field contours used to clear \pbar and \pos before trapped \Hbar are detected. The trap axis is vertical.}
\label{fig:HbarTrappingApparatus}
\end{figure}
}

\newcommand{\IoffeTrapQuenchFigure}{
\begin{figure}[htbp!]
\includegraphics*[width=\columnwidth]{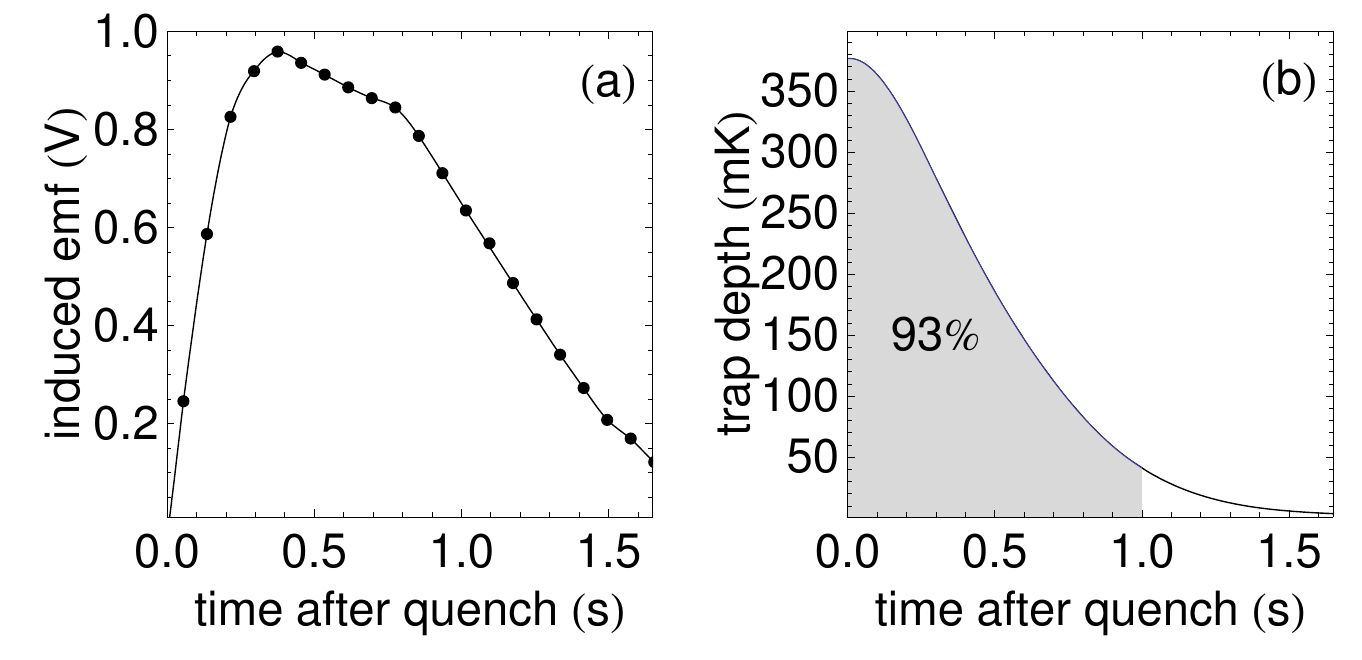}
\caption{(a) The emf induced across the field-boosting solenoid the Ioffe racetrack coil quenches.  (b) Deduced Ioffe trap well depth for ground state \Hbar shows if these fill the trap uniformly in energy that 93\% will escape within 1 s after the quench.}
\label{fig:IoffeTrapQuench}
\end{figure}
}

\newcommand{\HbarTimeSignalFigure}{
\begin{figure}[htbp!]
\includegraphics*[width=\columnwidth]{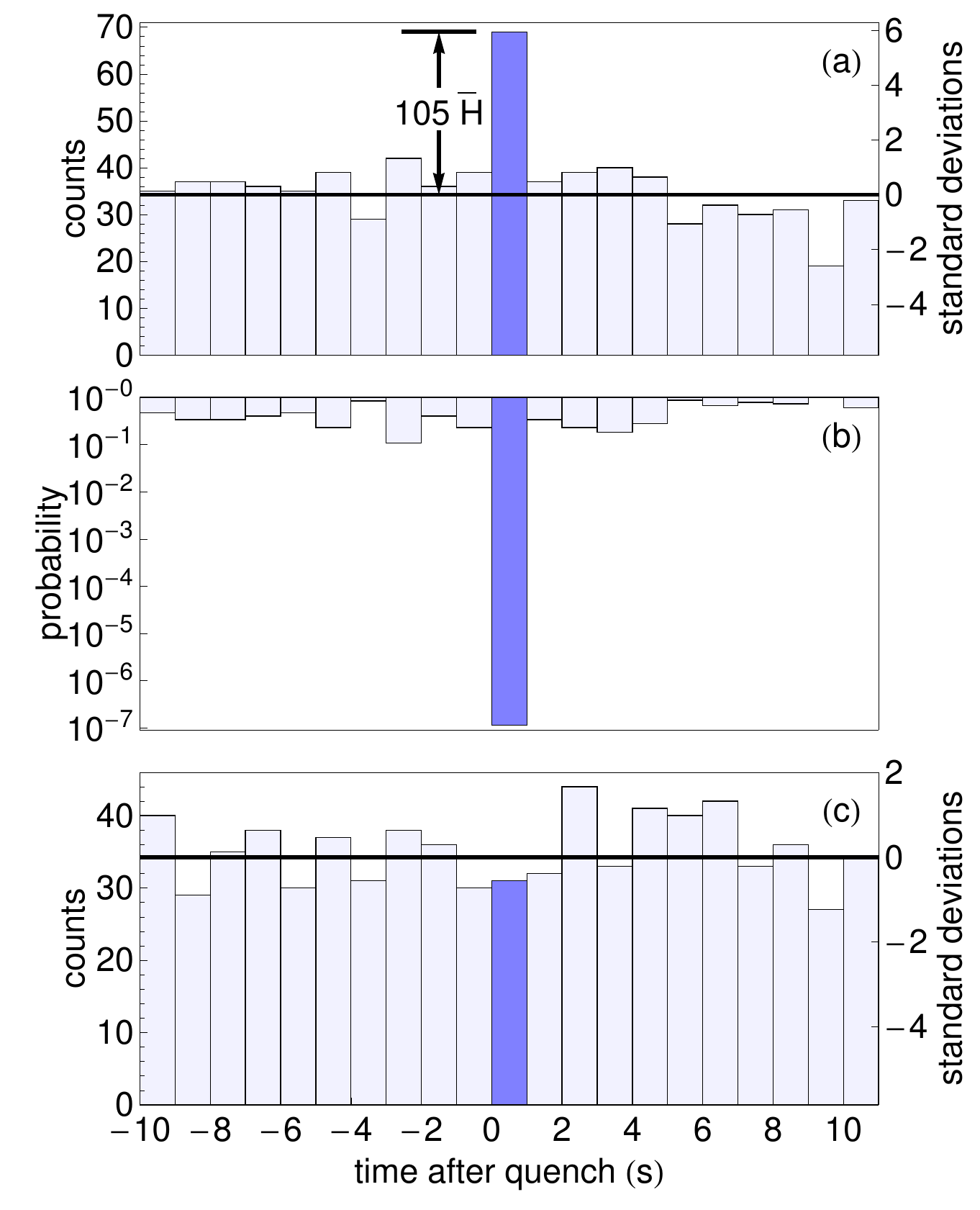}
\caption{(a) Detector counts in 1 s intervals for 20 trials.  The radial Ioffe trap field turns off and releases trapped \Hbar between $t=0$ and $1$ s.
 The counts in this interval above the average cosmic ray counts (solid line) correspond to 105 trapped \pbar for our detection efficiency.      (b) Probability that cosmic rays produce the observed counts or more.   (c) Quenching the Ioffe trap generates no false signals in 20 control trials.}
\label{fig:HbarTimeSignal}
\end{figure}
}

\title{Trapped Antihydrogen in Its Ground State}

\author{G.\ Gabrielse}\email[Spokesperson: ]{gabrielse@physics.harvard.edu}
\author{R.\ Kalra}
\author{W.S.\ Kolthammer}
\author{R.\ McConnell}
\author{P.\ Richerme}
\Harvard

\author{D.\ Grzonka}
\author{W.\ Oelert}
\author{T.\ Sefzick}
\author{M. Zielinski}
\Juelich

\author{D.W.\ Fitzakerley}
\author{M.C.\ George}
\author{E.A.\ Hessels}
\author{C.H.\ Storry}
\author{M.\ Weel}\York

\author{A.\ M\"ullers}
\author{J.\ Walz}\Mainz

\collaboration{ATRAP Collaboration}\noaffiliation

\date{12-27-2011 submitted to Physical Review Letters}

\begin{abstract}     
       Antihydrogen atoms (\Hbar ) are confined in an Ioffe trap for 15 to 1000 seconds -- long enough to ensure that they reach their ground state.  Though reproducibility challenges remain in making large numbers of cold antiprotons (\pbar) and positrons (\pos) interact, $5 \pm 1$  simultaneously-confined ground state atoms are produced and observed on average, substantially more than previously reported.  Increases in the number of simultaneously trapped \Hbar are critical if laser-cooling of trapped \Hbar is to be demonstrated, and spectroscopic studies at interesting levels of precision are to be carried out.
\end{abstract}


\maketitle

The original proposal to use cold \pbar and \pos to produce \Hbar atoms that could be confined in a magnetic trap for precise spectroscopy \cite{Erice} and gravitational measurements \cite{GabrielseGravity} is being actively pursued. Such spectroscopy (demonstrated later with \H \cite{KleppnerSpectroscopy}) could compare \Hbar and \H at a higher precision than the most stringent CPT tests with leptons \cite{DehmeltMagneticMoment} and baryons \cite{FinalPbarMass}. Cold \Hbar production \cite{2002AthenaNatureLetter,2002AtrapBackgroundFreeAntihydrogen}, at rates increased by driving forces \cite{2002AtrapDrivenAntihydrogenProduction}, led recently to a demonstration of almost one \Hbar atom per trial trapped for many seconds \cite{Alpha1000Seconds}.  \Hbar spectroscopy may eventually use only one trapped atom, but attaining interesting levels of precision will initially require many more simultaneously trapped atoms \cite{AntihydrogenSpectroscopy}.

When ATRAP produced the first \Hbar atoms in an Ioffe trap designed to confine them \cite{AntihydrogenProducedInPenningIoffeTrap}, no trapped atoms were observed. In an average of $N$ trials, ATRAP's detector could detect an average of $12/\sqrt{N}$ simultaneously trapped \Hbar atoms per trial at a $3\sigma$ level.  Increases in the detection efficiency, the number of trials, and/or the number of simultaneously trapped \Hbar were thus required.  The latter is the most attractive since more  simultaneously trapped \Hbar are needed to demonstrate laser cooling, and then precise laser and microwave spectroscopy.

To increase the likelihood that cold \Hbar would be produced, we cooled  the electrodes of the traps containing \pbar and \pos to 1.2 K \cite{PumpedHeliumRefrigerator}, demonstrated that \pos or \e stored within these thermalized at 20 K \cite{AtrapAdibaticCooling}, and prepared up to $10^7$ cold \pbar for  producing cold \Hbar \cite{AtrapAdibaticCooling}.  This Letter reports using $10^6$ \pbar (over an order of magnitude more than used for any previous attempt to trap \Hbar) to produce $5 \pm 1$ simultaneously trapped \Hbar atoms on average.  The \Hbar energies are below $375$ mK (with the low energy expressed in temperature units), and confinement times between $15$ and $1000$ s ensure that they are in their ground state.  The number of confined \Hbar produced   compares favorably to a very recent report \cite{Alpha1000Seconds} of $0.7 \pm 0.3$ atoms, with energies below $500$ mK stored from $50$ to $2000$ s, produced using many fewer \pbar.  Our demonstration that more \pbar produce more trapped \Hbar suggests that it may be possible to further scale up the number of simultaneously trapped \Hbar using the $10^7$ \pbar and many more \pos currently available.

\BtrapFigure

Each of the 20 trials in this demonstration take up to 2 hours.  One hour is used to accumulate and cool \pbar and 30 min.\ to accumulate and cool \pos. Up to 30 min.\ is used to bring the \pbar and \pos into position, ramp up the Ioffe trap, form \Hbar atoms, and look for trapped \Hbar.

Similar methods accumulate \pbar and \pos for all \Hbar experiments \cite{2005HbarReview}.  Within a $B=3.7$ T magnetic field, we accumulate $10^6$ \pbar from  6 to 8 pulses of $3 \times 10^7$ \pbar delivered with a 5 MeV energy approximately every 100 s by CERN's Antiproton Decelerator (AD).  The \pbar slow in a thin Be degrader window, are trapped within cylindrical Penning trap electrodes (Fig.~\ref{fig:Btrap}), and thermalize via collisions with trapped \e that are then ejected. The \pos from a $^{22}$Na source are trapped after they collide with gas molecules [11], and are
transported though a 9.5 m magnetic guide to enter the trap (from the right in Fig.~\ref{fig:Btrap}).
The trap electrodes are biased to form a nested Penning trap \cite{NestedPenningTrap} (e.g.\ Fig.~\ref{fig:HbarTrappingApparatus}a).  Approximately $3\times 10^7$ thermalized \pos   are positioned in the trap center, with $10^6$ \pbar at the bottom of the well to the right. After $B$ is reduced to 1 T, currents approaching  100 A are introduced into the pinch and racetrack coils \cite{AntihydrogenProducedInPenningIoffeTrap}.  This creates a 375 mK Ioffe trap for an \Hbar atom in its ground state (with equipotentials in Fig.~\ref{fig:HbarTrappingApparatus}b) to confine low-field-seeking \Hbar formed with sufficiently low kinetic energy.

\Hbar atoms form via a three-body interaction of a \pbar and two \pos \cite{NestedPenningTrap,GlinskyONeil}.  In a search for the most efficient production of cold \Hbar, the 20 trials differ primarily in the driving force applied to make the \pbar gain enough energy to pass through the \pos \cite{2002AtrapDrivenAntihydrogenProduction}.
To maintain resonance with the anharmonic \pbar center-of-mass oscillation as the \pbar oscillation energy increases, some trials apply a driving force with a frequency spectrum broadened by noise for up to 10 min.  Other  trials use a coherent drive, chirped in frequency for a duration of $2$ ms to $15$ min.  (See Fig.\ 3 of \cite{2002AtrapDrivenAntihydrogenProduction} and \cite{2011AlphaChirpedDrive}).

\TrapAndPotentialsFigure

\Hbar production and trapping continues for the 2 ms to 15 min.\ that the \pbar and \pos interact in the various trials. An \Hbar atom stays confined as long as its radiative decay takes it to another low-field-seeking state whose kinetic energy is less than the Ioffe trap well depth for the state. The \pbar and \pos are then cleared out by axial electric fields of about $\pm 5$ V/cm (Fig.~\ref{fig:HbarTrappingApparatus}a and c).  These fields, parallel to the trap axis, are chosen to be much stronger than any stray fields that  could otherwise trap a \pbar given the B field. For a \pbar to be trapped directly by the Ioffe trap, its cyclotron magnetic moment would require a 140 eV cyclotron energy -- much larger than the axial energy that could be transferred to cyclotron motion by \pbar-\pbar collisions.

The \Hbar remaining in the trap are released by quenching the superconducting racetrack coils of the Ioffe trap. The quench is caused by a heat pulse from a resistor (early trials) or by exceeding the critical current (later trials).

The minimum \Hbar storage time ranges from  $15$ to $60$ s (the time between the application of the clearing electric field and the quench).  However, in many trials the \Hbar storage time could be as long as the $1000$ s between when \Hbar production starts and the quench. The integral (Fig.~\ref{fig:IoffeTrapQuench}b) of the electromotive force (emf) induced (Fig.~\ref{fig:IoffeTrapQuench}a) in the field-boosting solenoid in Fig.~\ref{fig:Btrap} identifies the $1$ s time interval during which the radial Ioffe trap well depth is reduced so that 93\% of a uniform distribution of \Hbar energies can escape the trap.  In this interval, the signal from escaping \Hbar annihilating on an electrodes competes with the cosmic ray background.

\IoffeTrapQuenchFigure

The \pbar annihilation pions (and cosmic rays) are detected using 4 layers of 3.8 mm scintillating fibers (2 straight and 2 helical in Fig.~\ref{fig:Btrap}).  Made of BICRON BCF-12, with a peak emission wavelength of 435 nm and an attenuation length of 2.7 m, these 784 fibers are located outside the trap vacuum enclosure and the dewar that cools it (not shown in Fig.~\ref{fig:Btrap}). A double layer of 24 large scintillating paddles 1 meter high surrounds the dewar for the 1 Tesla superconducting solenoid (on a 66 cm radius, outside the view of Fig.~\ref{fig:Btrap}).  Coincidences between the fibers and paddles detect \pbar annihilations with an efficiency of 54\%, and a cosmic ray background of 41 Hz.

 A time-stamped record of the fibers and paddles triggered near the time of the quench is acquired at a rate up to $10^3$ events per second -- much higher than the observed count rate.  The probabilities that  4096 scintillator combinations are triggered by   \pbar annihilations and by cosmic ray are measured using the annihilations of $3 \times 10^5$ \pbar spilled radially and $3.5 \times 10^5$ cosmic ray events.  A Monte Carlo simulation shows that selecting the 256 scintillator combinations for which the ratio of these probabilities are greater than 4.55 optimize the signal-to-noise ratio -- providing enough signal while minimizing the effect of background fluctuations.  This choice reduces the background to 1.7 Hz from 41 Hz while decreasing the \pbar detection efficiency from 54\% to 33\%.

As described, the method used to make large \pbar and \pos plasmas interact is varied from trial to trial.  No clear favorite has yet emerged, and the interaction of the plasmas varies noticeably even for trials intended to be identical.  Accordingly, we sum the detector counts for all of the 20 trials carried out during the 2011 AD run.  The background averages down to allow a small signal from trapped \Hbar to become visible, suggesting that that some or all of the methods produce trapped \Hbar.

Fig.~\ref{fig:HbarTimeSignal}a shows the sum of the detector counts for the 20 trials in 1 s intervals, including the interval in which quenches emptied the Ioffe trap (colored blue).  The pronounced peak, when divided by the detector efficiency, shows that $105 \pm 21$ \Hbar atoms were trapped in the 375 mK quadrupole Ioffe trap.  This corresponds to an average of $5 \pm 1$ simultaneously trapped \Hbar per trial, stored in the trap for between $15$ and $1000$ s.  This signal is 6 standard deviations above what is expected from the observed background (right vertical scale in Fig.~\ref{fig:HbarTimeSignal}a), indicating that there is only 1 chance in $10^7$ that the signal in the central channel is a fluctuation of the cosmic background.  The counts in the 1 s intervals before and after the central signal interval are consistent with the measured statistical background.

Fig.~\ref{fig:HbarTimeSignal}c shows the sum of 20 control trials made by quenching the Ioffe trap when no \Hbar are trapped.  It shows that the sudden flux change from quenching the trap does not induce false coincidence signals that could be misinterpreted as being from \Hbar atoms.

The best of the 20 \Hbar trials illustrates current challenges and future possibilities. The count average and fluctuations outside of the central 1 s time interval are consistent with the other trials.   However, the counts in the central bin (corresponding to $39\pm8$ \Hbar atoms when the detection efficiency is included) are much higher than the average.  Sometimes we produce more \Hbar atoms and sometimes fewer, owing to our inability to precisely control the interaction of the \pbar and \pos, even in ``identical'' trials.  ``Identical'' trials also produce slightly different \pbar and \pos plasma diameters, and differences in the rate at which \pbar and \pos escape the nested trap.   If we analyze our trials without the best one (though we have no justification for discarding it) the average number of simultaneously trapped \Hbar per trial is $3.5 \pm 0.7$.  This is consistent with the average for all 20 trials, with a statistical significance of $4\sigma$ (a probability of less than $3 \times 10^{-4}$ of this being a background fluctuation).    Better control of the interaction of the large \pbar and \pos plasmas in a substantial magnetic gradient should produce the large number of trapped \Hbar in every trial.

\HbarTimeSignalFigure

To realize the long-time goal of precise \Hbar spectroscopy \cite{Erice} the \Hbar atoms must be in their ground state. The challenge is that ATRAP's field ionization method \cite{2002AtrapBackgroundFreeAntihydrogen} established that essentially all the \Hbar produced as \pbar and \pos interact in a nested Penning trap are in highly excited Rydberg states.  These highly excited, guiding-center atoms \cite{GlinskyONeil} are high-field-seeking states that cannot be trapped.

The trapped atoms must come from the small \Hbar fraction that the earlier field ionization measurements showed were produced with radii smaller than 0.14$ \mu$m \cite{2005HbarReview}. (This corresponds to a principal quantum number $n \approx 50$, though $n$ is not a very good quantum number for large $B$.)  Such atoms were shown to have chaotic \pos orbits \cite{2005HbarReview} owing to comparable strengths of the nonlinear Coulomb and magnetic forces on the \pos.  Some are apparently weak-field-seeking states, trapped via diamagnetic forces that are large for large $B$.

To remain in the trap for 15 to 1000 s, the radiating \Hbar  must remain in low-field-seeking states.  Simulations  \cite{PohlDecayInMagneticTrap,2006RobicheauxTBR} suggest this can happen, presumably because the angular momentum quantum number $m$ does not change on average, and $\Delta m = \pm1$ in a single radiative decay.

An overestimate of the time required for an $n=50$ state to decay to $n=1$ is the slowest radiation path, that from one circular state ($m = l = n-1$) to another.  The radiation rate for these states goes as $1/n^{5}$ and the $n=50$ circular state has a 30 ms lifetime. Rate equations describe a cascade to the ground state that takes about 0.5 s.  The actual cascade time is shorter given that fields and collisions mix in states with lower $l$ quantum numbers that radiate much more rapidly than circular states.  Thus  \Hbar detected after a $15$ to $1000$ s storage time are in their ground state.

Trapped \Hbar make it possible to compare their gravitational force $\kappa M g$ to the familiar $Mg$ on an H atom \cite{GabrielseGravity}.
Atoms created at the magnetic minimum on axis acquire $\kappa Mgh$ of energy in free fall to the magnetic maximum of the trap, with $h=10.6$ cm.
The atoms will escape a magnetic trap with an energy depth $W$ (375 mK in temperature units here) unless $|\kappa| \le W/(Mgh) = 3000$. For our trials, a $2\sigma$ level signal is present during the time that the radial well depth is reduced from 375 to 350 mK, establishing that $|\kappa| < 200$.  Improved limits will be possible with more trapped \Hbar, laser cooling, and probing of the \Hbar spatial distribution \cite{GabrielseGravity}.  It may eventually be possible to exceed the limit $|\kappa - 1| <1 \times 10^{-6}$ set by the consistency to better than 1 part in $10^{10}$ of \pbar and \p cyclotron clocks  \cite{FinalPbarMass}.  Such clocks would have different gravitational red shifts if the gravitation force differs by a factor of $\kappa$ for \pbar and p \cite{PbarGravity}.

In conclusion, more simultaneously trapped \Hbar atoms can be formed when large \pbar and \pos plasmas are used, despite the ongoing difficulties of controlling the interaction of large plasmas in a stable and reproducible way.  The approximately $105$ \Hbar atoms observed to be trapped in a $375$ mK, quadrupole Ioffe trap correspond to an average of $5$ simultaneously trapped \Hbar atoms per trial.  We are optimistic that increases in the number of trapped \Hbar atoms are coming.  Progress in manipulating and controlling the large and cold \pbar plasmas seems likely to continue. Lower plasma temperatures seem feasible. A new Ioffe trap just being assembled should make it possible to make the \pbar and \pos interact more efficiently. Even more simultaneously trapped \Hbar should be possible if these methods can be adapted for use with the ten times more \pbar that we have shown can be accumulated, and the ELENA upgrade to CERN's AD should make larger numbers of \pbar available.

We are grateful to CERN for providing 5-MeV \pbar and some support for W.O.  The NSF and AFOSR of the US, the BMBF, DFG, and DAAD of Germany, and the NSERC, CRC, and CFI of Canada support this work.

\newcommand{\Desktop}{\bibliography{d:/Jerry/Shared/Synchronize/ggrefs2011}}
\newcommand{\Laptop}{\bibliography{c:/Users/gabrielse/Jerry/Shared/Synchronize/ggrefs2011}}

%

\end{document}